\documentstyle[prl,aps,multicol]{revtex}

\def\crpar{\mbox{$\not\!\partial$}}
\def\crA{\mbox{$\not\!\! A$}}

\begin{document}

\title{Geometrical phases and quantum numbers of solitons
        in nonlinear sigma-models}
\author{A. G. Abanov$^1$ and P. B. Wiegmann$^2$}
\address{$^1$Department of Physics \& Astronomy,
SUNY @ Stony Brook, Stony Brook, NY 11794-3800, USA \\
$^2$James Franck Institute and Enrico Fermi
Institute
of the University of Chicago, \\
5640 S.Ellis Avenue, Chicago, IL 60637, USA
and
Landau Institute for Theoretical Physics}

\date{\today}
\maketitle

\begin{abstract}
Solitons of a nonlinear field interacting with fermions often acquire
a fermionic number or an electric charge if fermions carry a charge. 
We show how the same mechanism (chiral anomaly) gives solitons
statistical and rotational properties of fermions.  These properties
are encoded in a geometrical phase, i.e., an imaginary part of a
Euclidian action for a nonlinear $\sigma$-model.  In the most
interesting cases the geometrical phase is non-perturbative and has a
form of an integer-valued theta-term.

\end{abstract}
\vspace{0.15cm}

\begin{multicols}{2}
\narrowtext


{\bf 1. Introduction.}
The idea that a soliton of a nonlinear field interacting
with fermions may acquire quantum numbers proportional to its
own topological charge has a long history.  The first step
is traditionally attributed to T. H. R Skyrme who forty
years ago\cite{Skyrme} suggested that baryons emerge as
solitons of a meson field.  In 1968 Finkelstein and
Rubinstein \cite{FR} showed that a soliton may be a fermion.
The phenomenon became widely recognized probably after 1976
seminal papers of Jackiw and Rebbi \cite{JR}, P. Hasenfratz
and G. `t Hooft \cite{HH}, and A. S. Goldhaber
\cite{Goldhaber} who showed that
a monopole interacting with Dirac fermions acquires spin 1/2
through a zero mode of Dirac equation.  In condensed matter
physics Fr\"ohlich already in 1954 knew that in 1D
electronic system kinks of phonon modes carry an electric
charge \cite{Frohlich-1954}.  Later the appearance of
soliton quantum numbers has been linked to anomalous
properties of current algebras\cite{TJZW-1985}.
The papers which shaped the
subject can be found in the reprint volume
\cite{RebbiSoliani}.
Nowadays the phenomena when topological spatial
configurations of a nonlinear bose field acquire fermionic
charges and through them rotational and statistical
properties of fermions are ubiquitous.

Fermionic number, spin, and statistics of solitons emerge
differently.  The first can be probed for by an external
gauge field.  The second and the third emerge through a
geometrical phase of the wave function or, equivalently,
through a topological term in an effective bosonic
action\cite{WilczekZee-1983,Polyakov}.  The geometrical
phase changes by $2\pi$ times a quantum number (spin or
statistics) in an adiabatic motion along a closed path in
the configurational space which corresponds to a full
rotation or to an interchange of solitons respectively.

Intuitively one expects that once a soliton acquires
fermionic charge it must inherit fermionic rotational
properties as well.  This is indeed true although a relation
between these two aspects remains relatively obscure.  In
the most interesting cases the geometrical phase is a
theta-term, i.e., an integer-valued topological invariant.  In
this case it does not appear in perturbation theory contrary
to the fermionic charge.  We address this issue in this
letter, using some results of our recent paper \cite{AW}.
For an earlier discussion see Ref.\cite{Jaroszewicz-1984}.

To simplify an analysis we concentrate on a particular
example of a sigma model in (2+1) dimensions. Generalization
to more general case is relatively straightforward but more
mathematically involved. We do not attempt it in this
publication (see, however, the conclusion).


{\bf  2. Fermionic $\sigma$-models.}
We consider fermionic $\sigma$-model\cite{applications}
\begin{equation}
     \label{100}
        \hspace{1.0cm} {\cal L}= \bar\psi \big(i\hat D
        +im\cdot\hat n\big)\psi.
\end{equation}%
Here $\hat{D}=\gamma_\mu(\partial_\mu-iA_\mu)$, where
$A_{\mu}$ is an external Abelian gauge field (e.g.,
electromagnetic field), $\gamma_{\mu}$ are Dirac
gamma-matrices, and $\hat{n}$ is a matrix field which
belongs to some compact target space $V$.  We set $\hat
n^2=1$.  We use a Euclidian formulation with
$\{\gamma_{\mu},\gamma_{\nu}\}=2\delta_{\mu\nu}$.  Fermions
are Dirac spinors in the Euclidian spacetime.  In addition
they also form an isomultiplet on which matrix $\hat{n}$ acts.

The nature of a geometrical phase depends on the topology of
a target space and of a spacetime.  In order to illustrate
how a soliton with a unit topological charge and a unit
fermionic charge becomes a relativistic particle with a spin
1/2 we choose a particular target space $V=CP^M$
\cite{Jaroszewicz-1984,Abanov} and a three-dimensional (2+1)
spacetime.

For any integer $M$ static configurations of $\hat n$ fall
into homotopy classes $\pi_{2}(CP^{M})=Z$.  This means that
among configurations of $\hat n$ there are topologically
non-trivial ones -- solitons.  The
$CP^M$- target space can be parameterized by a complex
$M+1$-component unit vector
$z=(z_{1},z_{2},\ldots,z_{M+1})^{t}$ with constraint
$z^{\dagger}z=1$.
Then the matrix $\hat n$ is $(M+1)\times(M+1)$ matrix
$\hat{n}=2zz^{\dagger}-1$ (so that $\hat{n}^{2}=1$).  It does
not change when $z$ is multiplied by an arbitrary phase
which means that the target space is $CP^M$.  In
the special case $M=1$, the target space becomes
two-dimensional sphere $CP^{1}=S^{2}$ and
$\hat{n}=\vec{n}\vec{\tau}$, where
$\vec{n}=(n_{1},n_{2},n_{3})=z^{\dagger}\vec{\tau}z$ is a
unit vector $\vec{n}^{2}=1$ and $\vec{\tau}$ is a set of
three Pauli matrices acting in isospace.

There is an important difference between $CP^{1}$ model and
$CP^{M}$ model with $M>1$.  The former has a symmetry with
respect to a  parity transformation $x_{1}\to -x_{1}$
supplemented by an internal transformation
$\vec  n\to -\vec n$.  Contrary, the
$CP^{M}$ model does not have this symmetry because the
change $\hat{n}\to -\hat{n}$ can not be
achieved by a transformation within the target space.  The
parity is explicitly broken  and is not a symmetry of the
problem for $M > 1$.


{\bf 3. Nonlinear $\sigma$-models.}
The nonlinear $\sigma$-model ${\cal W}[n]$ is a result of
an integration over fermions
\begin{equation}
            \label{11}
              e^{-{\cal W}}
              = \int D\psi D\bar\psi\, e^{-\int dx{\cal L}}
              =\mbox{Det}\big(i\hat D  +im\cdot\hat n\big).
\end{equation}
For smooth and slow configurations of $n$-field the
effective action is local and can be systematically
studied in $1/m\;$ (gradient) expansion.  In a Euclidian
spacetime it has an imaginary part ${\cal W}= Re{\cal W} +
Im {\cal W}$.  The leading terms of the gradient expansion
of a nonlinear $\sigma$-model (we compute them later) are
\begin{eqnarray}
     Re\,  {\cal W} &=& \frac{m}{16\pi}\int d^{3}x\;
     \mbox{tr}\,\, (\partial_\mu \hat n)^2,
  \\
     Im\, {\cal W} &=& -i\pi\,  \Gamma - i\int d^{3}x\;
     A_{\mu} J_{\mu},
  \label{89}
\end{eqnarray}
where the ``topological current'' $J_{\mu}$ is defined as
\begin{equation}
  \label{eq:topcurrd3}
     J_{\mu} = -\frac{i}{16\pi}\epsilon_{\mu\nu\lambda}\,
     \mbox{tr}\,\left(\hat n \partial_{\nu}\hat n
     \partial_{\lambda}\hat{n}\right)
\end{equation}
and $\pi\,\Gamma$ stays for the   geometrical phase.

The two terms of an imaginary part (\ref{89}) of the effective action
reflect different quantum numbers of solitons.  They are of
the topological origin and remain in the action even in the
limit of infinite mass $m\to \infty$.  The second term in
(\ref{89}) is a
response to an external field.  It establishes a
relation\cite{GoldstoneWilczek-1981,Jaroszewicz-1984,NiemiSemenoff-1986}
between the fermionic current in the background $n$-field
$j_{\mu}=i\langle \bar{\psi}\gamma_{\mu}\psi\rangle$ and the
topological current $J_{\mu}$
\begin{equation}
              \label{J}
                j_{\mu}=i\frac{\delta}{\delta A_{\mu}}{\cal W}
                =J_\mu +O(1/m).
\end{equation}
It represents a well-known fact that a soliton acquires
fermionic charge due to a chiral anomaly.  The total number
of fermionic levels with positive energy changes by $Q$ if
the topological charge of the vacuum changes by $Q$
\begin{equation}
              \label{J1}
                 \int_{\rm space}d^{2}r\; j_{0}=Q
	\equiv \int_{\rm space}d^{2}r\; J_{0}.
\end{equation}
Note that although eq.~(\ref{J}) has higher gradient
corrections, eq.~(\ref{J1}) is exact: a soliton with a
topological charge $Q$ carries fermionic number $Q$.

Let us assume that the coordinate space is
compactified into a two-dimensional sphere $S^{2}$.  Then a
smooth static configuration of bosonic field $\hat{n}(r)$
defines a mapping of $S^{2}$ into the target space
$V=CP^{M}$.  All mappings $S^{2}\to CP^{M}$ can be divided
into distinct topological (homotopy) classes so that
any two configurations within one class can be smoothly deformed one
into another.  Classes are enumerated by an integer number
$Q$ -- the topological charge, which in this case can be any integer
number.  The integer $Q$ can be expressed as a spatial
integral of the density of topological charge $Q=\int
d^{2}r\;J_{0}(r)$ where $J_{0}$ is a zeroth component of
the conserved topological current (\ref{eq:topcurrd3})
\cite{homology}.

Because the topological current is local, one can think
about localized topological solitons.  In particular a
special configuration with zero net topological charge which
consists of two well-separated positive and negative bumps
of unit topological charge can be considered as a separated
soliton-antisoliton pair.

The first term in (\ref{89}) is a geometrical phase.  It
represents the change of the phase of a fermionic
wavefunction due to a change of the background configuration
of $\hat{n}$. The following spacetime configurations are of
special interest. The first one is a process of a spatial
$2\pi$-rotation of a single localized soliton around some
axis. Another one is an interchange of two
solitons\cite{processes}.  If a soliton has spin $S$ and
$\theta$-statistics, the phase of the fermionic wavefunction
(the geometrical phase) changes in these processes by $2\pi S$
and $\theta$ respectively, where $\theta=\pi,0$ for a
fermion (boson).  It is natural to expect that since soliton
acquired a unit of a fermionic charge, it should also
acquire rotational numbers of a fermion, i.e., spin $S=1/2$
and fermionic statistics $\theta=\pi$.  This
corresponds to the change of the phase $\pi\,\Gamma$ by
$\pi$ for both processes.  We will show
on the example of model (\ref{100}) that this is indeed
the case.


{\bf 4.  Calculation of fermionic number.} We start
from the calculation of the fermionic number of a
soliton -- the second term in (\ref{89}). This computation is
perturbative
\cite{GoldstoneWilczek-1981,Jaroszewicz-1984,NiemiSemenoff-1986}
and constitutes the chiral anomaly.  We sketch it below.

A variation of an effective action with respect to a gauge
field $A_{\mu}$ is $\delta_{A}{\cal W}=
-\mbox{Tr}\,\delta_{A}D D^{-1} =-\mbox{Tr}\left\{\delta \crA
{\cal D}^{\dagger}({\cal D}{\cal D}^{\dagger})^{-1}
\right\}$, where ${\cal D}=i\hat D +i m \hat n$, ${\cal
D}^{\dagger}=i\hat D -i m \hat n$, and ${\cal D}{\cal
D}^{\dagger} =(i\crpar+\crA)^{2} +m^{2}+m\crpar \hat n $.
Expanding the denominator in gradients of chiral fields and
keeping only terms linear in $A$ \cite{CSterms} in the
imaginary part of the action we obtain
\begin{equation}
  \label{00}
     \delta_{A}Im {\cal W} = \mbox{Tr}
     \left\{\delta\crA {\hat n}\frac{im}{-\partial^2+m^2}
     \left(\crpar {\hat n}
     \frac{m}{-\partial^2+m^2}\right)^{2}
     \right\}.
\end{equation}
Calculating the functional trace and the trace over $2\times
2$ gamma-matrices, we obtain eq.~(\ref{J}), i.e., the
fermionic current is equal to the topological current
(\ref{eq:topcurrd3}).  This is the first nonvanishing term
in the gradient expansion for the current.


{\bf 5.  Geometrical phase and topology of target space.}
Spacetime configurations of chiral bosonic field
$\hat{n}(r,t)$ define mappings of a spacetime into the
target space $V$.  Here we consider the spacetime
compactified into a sphere $S^{3}$ so that we
have mappings $S^{3}\to V$\cite{torus}.
We would like to distinguish
between two possibilities.  The first is when $\pi_{3}(V) =
0$, i.e., any two of spacetime configurations (mappings) can
be smoothly deformed one into another.  In this case the
geometrical phase is perturbative.  The second possibility
is when $\pi_{3}(V)=Z\neq 0$ and configurations are divided
into topological (homotopy) classes.  The $CP^{M}$ model in
(2+1) realizes both cases: $\pi_{3}(CP^{M})=0$ at $M>1$ and
$\pi_{3}(CP^{1})=\pi_{3}(S^{2})=Z$.  In the latter case
noncontractable spacetime configurations are labeled by the
Hopf invariant.

To calculate the geometrical phase we
put $A_{\mu}$ to zero and calculate
a variation of an effective
action over the chiral field $\hat n$
\begin{eqnarray}
     \delta Im\,{\cal W} &=& -\Im m\,\mbox{Tr}\,
     (im \delta \hat n {\cal D}^{\dagger}
     ({\cal D}{\cal D}^{\dagger})^{-1} )
  \label{v} \\
     &=& \mbox{Tr}\left\{m
     \delta {\hat n}{\hat n} \frac{m}{-\partial^2+m^2}
     \left(\crpar {\hat n}
     \frac{m}{-\partial^2+m^2} \right)^{3} \right\}
  \nonumber \\
     &=&\frac{i}{32\pi} \int_{S^{3}}d^{3}x\, \epsilon_{\mu\nu\lambda}
     \mbox{tr}\left(\delta \hat n {\hat n}
     \partial_{\mu} \hat n \partial_{\nu}
     \hat n \partial_{\lambda} \hat n\right).
  \nonumber
\end{eqnarray}
First, consider the case when all spacetime configurations
of $\hat{n}$ are contractable: $\pi_{3}(V)=0$.  In this case
one can restore the geometrical part of the action from its
variation (\ref{v}).  For $V=CP^{M}$ we write $\hat n =
2zz^{\dagger}-1$ and (after some algebra) we obtain for $M\neq 1$
\begin{eqnarray}
    \delta Im\,{\cal W}
    &=& \frac{i}{4\pi} \int_{S^{3}}d^{3}x\,
    \delta (\epsilon_{\mu\nu\lambda}
    a_{\mu}\partial_{\nu}a_{\lambda}),
  \label{eq:resGam}
\end{eqnarray}
where a gauge field is $a_{\mu}\equiv
-iz^{\dagger}\partial_{\mu} z$.  This gives the geometrical
phase
\begin{eqnarray}
       \Gamma &=&-\frac{1}{4\pi^{2}} \int_{S^{3}}d^{3}x\,
       \epsilon_{\mu\nu\lambda}
        a_{\mu}\partial_{\nu}a_{\lambda}.
     \label{eq:Gamma}
\end{eqnarray}
In this case the geometrical phase is perturbative. Under
parity transformation the geometrical phase (\ref{eq:Gamma})
changes sign. It reflects the fact that the model (\ref{100}) is not
invariant under parity transformation.

This method does not work, however, in the most interesting
case when $\pi_{3}(V)\neq 0$.  In this case the variation of
the  geometrical phase over chiral field is identically
zero.  E.g., for $CP^{1}$ model
$\epsilon_{\mu\nu\lambda}\mbox{tr}\left(\delta \hat n {\hat
n} \partial_{\mu} \hat n \partial_{\nu} \hat n
\partial_{\lambda} \hat n\right)=0$.  This does not mean,
however, that the geometrical phase vanishes.  It vanishes
perturbatively, but in fact it is proportional to a
topological invariant labeling the class of a mapping from
$\pi_{3}(V)$.

The strategy to go around the topological obstacle has been
suggested by
Witten\cite{Witten-1983baryons,Witten-1983global}.  It is
known as a method of embedding\cite{ElitzurNair-1984}.  The
idea is to extend the fermionic $\sigma$-model with the
target space $V$ to a one with a bigger space $\tilde V$, so
that any mapping of a spacetime into $\tilde V$ is
topologically trivial.  Then the geometrical phase varies
under the variation of $\hat n$ so that (\ref{v}) does not
vanish and one can use the perturbative result (\ref{v}) to
restore the geometrical phase.  Then the proper reduction to
the target space of interest $\tilde V\to V$ should be made
in such a fashion that (\ref{100}) becomes the model of
interest plus decoupled free fermions.  Under this reduction
the geometrical phase becomes  an integer-valued topological
invariant.  This method was applied to $CP^{1}$
model\cite{Abanov} with an extension to a topologically
trivial $\tilde{V}=CP^{M}$ model.  The result for the
geometrical phase is given by the same eq.~(\ref{eq:resGam})
as for $M>1$.  For $M=1$, however, it is an integer-valued Hopf
invariant\cite{Monastyrskii}.
The geometrical phase in this case is $\mbox{mod}(2\pi)$
invariant under parity transformation as well as fermionic
model (\ref{100}) at $M=1$.

To calculate the geometrical phase we compare the effective
action for two physically different configurations of chiral
field, e.g., corresponding to a soliton at rest and to a rotating
soliton.  These configurations belong to
different topological classes and, therefore, it is
impossible to interpolate between them smoothly.  Singular
interpolating configurations require an ultraviolet regularization of
the theory.  The embedding method essentially provides us with such a
regularization.  It allows to connect topologically
different configurations by a smooth interpolation through
an extended target space\cite{globalanomaly}.

Some  consequences of  the geometrical phase in the model with
$M=1$ have been discussed in  \cite{AW1}.


{\bf 6.  Relation between topological current and
geometrical phase.} Now we are ready to show that under a
rotation of a localized soliton or when interchanging positions of two
solitons the geometrical phase changes by $\pi$ in accord
with the unit fermionic charge of a soliton.  This
establishes a  charge-spin-statistics correspondence: a soliton
with a topological charge $Q$ possesses the fermionic number
Q, the spin Q/2 and the fermionic (bosonic) statistics for
$Q$ odd (even) respectively.  Let us consider a soliton
slowly rotating  by an angle $2\pi$ and compare
the geometrical phase of this process with the one without
rotation.  For simplicity we choose a static localized
soliton $\hat n(r)$ with a spherically symmetric topological
charge density $J_0(r)$.  Topological current is related to
a gauge field $a_{\mu}$ by
$J_{\mu} = \frac{1}{2\pi} \epsilon_{\mu\nu\lambda}
\partial_{\nu}a_{\lambda}$. We choose a vector potential
$a_{\mu}=(a_{0},\vec{a})$, such that $a_{0}=0$, and
$\vec{\nabla}\times \vec{a} =2\pi J_0(r)$.  Then the vector
potential $a_{\mu}=(\omega(t) (q(r)-1), \vec{a})$ with
$q(r)=\int_{0}^{r}dr' 2\pi r' J_0(r')$ corresponds to a
soliton rotating with the angular velocity $\omega(t)$.  Its
topological current is
$J_{\mu}=(J_0(r),J_0(r)\omega\epsilon^{ab}r_{b})$.  The
geometrical phase (\ref{eq:Gamma}) of the rotating soliton
is $-\frac{1}{4\pi^{2}}\int d^{2}r\,
2a_{0}\vec{\nabla}\times\vec{a} = \frac{\omega}{2\pi}Q$ which
gives $\Gamma =Q\int dt\,
\frac{\omega(t)}{2\pi}=Q\frac{\Delta\phi}{2\pi}$.  A
$2\pi$-rotation of soliton gives rise to the phase
$\pi\Gamma=\pi Q$.  This corresponds to the spin $Q/2$ of
the soliton.  Similar arguments can be applied for the
process of an interchange of positions of two solitons.  The
result is the same for $CP^{M}$ field with any $M$ and
remains valid for $M=1$, when the geometrical phase is $\pi$
times an integer Hopf number.


{\bf 7.  Conclusion.} To summarize, we argued that as soon
as a soliton of a nonlinear field acquires a fermionic
number as a result of an interaction with fermions, the
soliton acquires also all other fermionic quantum numbers:
statistics and spin.  These quantum numbers are represented
by a geometrical phase in an effective action of the
nonlinear $\sigma$-model.  In some interesting cases when
the configuration space has noncontractable paths, the
geometrical phase becomes an integer-valued $\theta$-term.
We discussed the both cases on the example of
$2+1$-dimensional theory where a single species of fermions
interacting with a chiral bose field taking values in $CP^{M}$
target space.  However, the fermionic number-spin-statistics
correspondence is a general phenomenon.  An extension of our
results to more general cases: arbitrary number of fermionic
species, other target spaces, and dimensions (both odd or
even) is relatively straightforward\cite{AW}.

In general, we formulate (in physical
terms) the following theorem (conjecture).
If (i) mappings of the
coordinate space into a target space of a nonlinear bose field
are noncontractable\cite{theorem} and (ii) a noncontractable
static configuration carries a fermionic number as a result
of an interaction with fermions, then the effective action
of the bose field possesses a geometrical phase which
changes by $\pi$ times fermionic number of a soliton under
$2\pi$-spatial rotation of a localized soliton.  If solitons
are localized the same phase appears in a process of
interchanging of positions of solitons.


{\bf Acknowledgement.}
The authors are grateful to  S.~P.~ Novikov and 
M. ~I.~Monastyrsky for discussing some 
aspects of topology.  A.G.A.
would like to acknowledge receipt of an Alfred P. Sloan
fellowship.  P.B.W. was supported by grants NSF DMR 9971332
and MRSEC NSF DMR 9808595.


\end{multicols}


\begin{references}
\bibitem{homology}
The mechanism discussed in the paper holds for a fermionic
model (\ref{100}) in d+1 dimensions with both non-trivial
$d$-homotopy group
$\pi_{d}(V)=Z$ and $d$-cohomology group
of the target space $H_{d}(V)=Z$.
The latter means that there exists a
topological current -- closed but not exact $d$-form.
\bibitem{CSterms}
We consider here only terms linear in the external gauge field
$A$ and omit the Abelian Chern-Simons term present in
$CP^{M}$ models (with $M>1$) because of an explicitly broken
parity.  This Abelian Chern-Simons term is important but
does not play any role in this letter.
\bibitem{torus}
A more physical compactification of spacetime is the one
into the product space $S^{2}\times S^{1}$.  In this case an
explicit expression for a geometrical phase is much more
involved.  In particular the
compactification into product spaces leads to some
restrictions on the allowed spin and statistics of solitons.
However, it does not change our conclusions.
\bibitem{theorem}
More rigorously one should require the non-triviality of
$H_{d}(V)$ and assume the conditions of Gurevich theorem to
avoid the multitude of lower-dimensional topological terms.
\end{references}
\end{document}